\documentclass[prd,showpacs,twocolumn]{revtex4}

\newcommand{\beq}{\begin{equation}}
\newcommand{\eeq}{\end{equation}}
\newcommand{\beqar}{\begin{eqnarray}}
\newcommand{\eeqar}{\end{eqnarray}}

\newcommand{\ket}[1]{| #1 \rangle}
\newcommand{\bra}[1]{\langle #1 |}
\newcommand{\inprod}[2]{\langle #1 | #2 \rangle}

\ifx\undefined\fiverm
     \font\fiverm=cmr5
\fi

\input{prepictex}
\input{pictex}
\input{postpictex}

\begin{document}
\author{Tarun Biswas}
\title{Nonunitary Interaction, Adiabatic Condition, Haag's Theorem and Renormalization}
\email{biswast@newpaltz.edu}
\affiliation{State University of New York at New Paltz, \\ New Paltz,  NY 12561, USA.}
\date{\today}
\begin{abstract}
Haag's theorem has shown that the transformation between interacting and free field operators
in a reasonable quantum field theory cannot be unitary. Here, the original requirement of unitarity is
revisited from a physical point of view to show not only that unitarity is not required but indeed
not possible. Electrodynamics is used as an example. In a classical treatment the interaction 
cannot be turned on or off adiabatically as energy conservation cannot be maintained in a
physically meaningful way. In a fully second quantized theory the interaction (or source) term
is always present in the equation of motion even if the system is in the vacuum state. So,
the interaction cannot be physically turned on or off adiabatically or otherwise. Hence, the transformation
$V(t)$ from free fields to interacting fields cannot be interpreted as an actual time evolution. This
makes the unitarity of $V(t)$ quite unnecessary as the original reason for unitarity was to
conserve probability (or energy) through a time evolution. It is conjectured that the infinite terms in the
scattering matrix that need renormalization appear due to the forcing of $V(t)$ to be unitary.
An alternate approach to scattering of interacting fields is suggested.
\end{abstract}
\maketitle
\section{Introduction}
Quantum field theories (QFTs) have been very successful in modeling various kinds of interactions
in particle physics. However, there are two notable problems in the general formulation
of these theories. First, infinite terms are found to appear in perturbative computations of interaction
processes. Renormalization procedures seem to handle these infinities in interesting ways.
However, infinite terms in perturbative expansions are still mathematically unacceptable.
Second, Haag's theorem\cite{wightman} has shown, quite unequivocally, that the transformation 
from free fields to interacting fields cannot be achieved by a unitary operator. It is very likely
that the two problems are related and solving one may solve the other automatically. A
possible scheme to achieve such a solution has been presented in the past\cite{biswas}.

Although Haag's theorem requires it, a nonunitary interaction operator $V(t)$ is a matter of significant
discomfort. This is because unitarity of $V(t)$ is the basis of most computations in field theory.
So, here, the necessity of a unitary $V(t)$ will be revisited. It is noted that the requirement of a
unitary $V(t)$ stems from its proposed relation to the time evolution operator $U(t)$ -- The effect
of $V(t)$ is seen as time evolution during a period over which the interaction is turned on and then
off (the adiabatic condition). The adiabatic condition has been used effectively in first
quantized scattering theories. However, in the following, it will be argued that this interpretation 
is incorrect for QFT. Hence, the requirement of a unitary $V(t)$ is unnecessary.
Besides, in some interpretations, the unitarity of a general time evolution 
$U(t)$ itself can be questioned in the realms of
quantum field theory. Nonunitary $U(t)$ has been used effectively in some applications of field
theory\cite{lamb} where the time evolution of a subset of the Hilbert space is studied. In the
present discussion, there may be some hint of such operations in subsets of the full Hilbert
space. However, for all practical purposes the unitarity of $U(t)$ will be maintained.

\section{The need for a unitary time evolution}
In first quantized quantum mechanics the time evolution of a state $\ket{\psi(t)}$  is
given by
\begin{equation}
\ket{\psi(t)} = U(t)\ket{\psi(0)},
\end{equation}
where $t$ is time and $U(t)$ is the time evolution operator. The Schr\"{o}dinger equation gives
\begin{equation}
U(t)=\exp(-iHt),
\end{equation}
where $H$ is the Hamiltonian. A hermitian $H$ results in a unitary $U(t)$ which results in a time
independent total probability $\inprod{\psi(t)}{\psi(t)}$. Hence, it is noted that the original
motivation for a unitary time evolution operator is conservation of particle number. This can be
generalized to the requirement of energy conservation. In a relativistically covariant theory, with 
no external fields, this leads to 4-momentum conservation. This motivation from 4-momentum
conservation remains in second quantized theories (QFT) as well. In such theories, the original
particle number conservation is no longer true but 4-momentum conservation still leads to
a unitary $U(t)$. A QFT for free fields is straightforward. However, the introduction of interactions of
fields makes things tricky. The identification of the Hilbert space of the free fields
to that of the interacting fields is itself questionable. However, the standard methods in
QFT make such an assumption. Standard methods also assume that an interaction 
operator $V(t)$ transforms free fields to interacting fields. Such a transformation is given the 
meaning of a time evolution by assuming that the interaction can be turned on and off over 
time (adiabatic condition) thus transforming free fields to interacting fields and back. This
interpretation of $V(t)$ as time evolution is at the root of the requirement that it be unitary.
Hence, we need to study the adiabatic condition more closely. We shall use electrodynamics
as an example.

\section{The adiabatic condition in classical electrodynamics}
Consider the source free equation for the classical electromagnetic 4-potential
$A^{\mu}_{f}$:
\begin{equation}
\partial_{\nu}\partial^{\nu}A^{\mu}_{f}(x)=0,
\end{equation}
where $x$ represents $x^{\mu}$ the 4-position. The subscript $f$ is used to denote
``source free''. In the presence of a charge current $j^{\mu}$, this equation becomes:
\begin{equation}
\partial_{\nu}\partial^{\nu}A^{\mu}(x)=-j^{\mu}(x).
\end{equation}
The general solution for this is
\begin{equation}
A^{\mu}(x)=A^{\mu}_{f}(x)+A^{\mu}_{s}(x),
\end{equation}
where the particular solution $A^{\mu}_{s}$ is given by:
\begin{equation}
A^{\mu}_{s}(x)=\int G(x-x')j^{\mu}(x')d^{4}x',
\end{equation}
and $G(x-x')$ is the Green function. In QFT the current $j^{\mu}$ is seen as the
interaction term. So, the adiabatic condition requires that it be possible to
turn on and turn off $j^{\mu}$ over time. However, if $j^{\mu}$ were turned on 
and then off, the energy of the system would change due to the change in
$A^{\mu}_{s}$. Hence, energy would not be
conserved over time. So, the adiabatic condition is fundamentally at odds with energy
conservation and consequently with the depiction of interaction as time evolution.

A related problem of turning $j^{\mu}$ on or off is seen in the following. For the adiabatic
condition, consider the
current to be turned on at a time $t=-T$ and turned off at a time $t=+T$. This is
represented by a current $j^{\mu}_{T}$ as follows.
\begin{equation}
j^{\mu}_{T}=g_{T}(t)j^{\mu},
\end{equation}
where
\begin{equation}
g_{T}(t)=\left\{\begin{array}{ll}
1, & \mbox{for } -T<t<T \\
0,  & \mbox{otherwise}
\end{array}\right .
\end{equation}
Charge conservation requires that
\begin{equation}
\partial_{\mu}j^{\mu}=0.
\end{equation}
However,
\begin{equation}
\partial_{\mu}j^{\mu}_{T}=g_{T}(\partial_{\mu}j^{\mu})+(\partial_{\mu}g_{T})j^{\mu}\neq 0.
\end{equation}
The time derivative of $g_{T}$ is zero everywhere except at $t=-T$ and
$t=+T$ where there are Dirac delta infinities. So charge conservation is
violated by the adiabatic condition. This is a problem even if $T\rightarrow\infty$
as it only moves the Dirac delta infinity to an infinity in time.

One way of saving the situation might be to separate out the part of $A^{\mu}_{s}$ that is
produced by $j^{\mu}$ in the absence of $A^{\mu}_{f}$. The rest of
$A^{\mu}_{s}$ would be due to changes in $j^{\mu}$ caused by $A^{\mu}_{f}$. 
These higher order effects may be expected to be small enough such that energy conservation
is at least approximately satisfied. For example, consider a stationary point charge
that starts oscillating due to an incident plane wave packet. If the energy of the
field of the static charge is subtracted from the total energy, one might still find
a meaningful perturbative solution. However, such a separation is not always
possible. In particular, in a QFT this becomes very difficult as charges can be
created and annihilated by the scattering process.
 
The adiabatic condition is often used successfully for first quantized systems. This is
because in such systems energy does not include rest-mass energy and nothing is
created or annihilated. A typical case is that of an electron colliding with another charge.
Here turning the interaction on or off results in particle energy conversion from kinetic
to potential and back with no net loss in energy. In a field theory, the interaction term has a 
somewhat different meaning. It has an energy of its own and hence, turning it on or off
adds (or subtracts) energy to the system.

\section{The adiabatic condition in quantum electrodynamics}
In quantum electrodynamics (QED) second quantization adds to the troubles of
the adiabatic condition. The absence of charge in a particular state of the system
does not imply $j^{\mu}=0$. Empty space means a vacuum state. It does not mean
that field operators themselves have zero values. Hence, in QED, the equations of
motion of the physical system are always the interacting field equations.
\begin{eqnarray}
\partial_{\nu}\partial^{\nu}A^{\mu} & = & -j^{\mu}, \\
(\gamma^{\mu}\partial_{\mu}+m)\psi & = & -ie\gamma^{\mu}A_{\mu}\psi, \\
j^{\mu} & = & -ie\bar\psi\gamma^{\mu}\psi,
\end{eqnarray}
where $A^{\mu}$ are the photon fields and $j^{\mu}$, the current,
is written in terms of the electron spinor fields $\psi$.

These field equations are relationships among field operators and they hold for
all states of the system -- even the vacuum state which has no observable particles.
Physically, turning the interaction on or off can, at best, be interpreted as adding
or removing particles -- in other words, changing the state of the system. It would not mean
adding or removing terms in the field equations. Hence, besides
the problems seen in classical electrodynamics, the adiabatic condition has no meaning
in QFT. 

\section{Relating time evolution and interaction operators}
The free field equations may have no physical meaning, but they can still be written 
down mathematically.
\begin{eqnarray}
\partial_{\nu}\partial^{\nu}A^{\mu}_{f} & = & 0, \\
(\gamma^{\mu}\partial_{\mu}+m)\psi_{f} & = & 0, 
\end{eqnarray}
where the subscript $f$ is used to denote the free fields in each case.
One may also define an operator $V(t)$ that transforms from free fields to
interacting fields. But, due to the meaninglessness of the adiabatic condition, it cannot
be interpreted as a time evolution. Hence, $V(t)$ need not be unitary. This can save
perturbative QFT from Haag's theorem. Let ${\cal H}$ be the Hilbert space of interacting
fields and ${\cal H}_{f}$ the Hilbert space of free fields\footnote{The reason for keeping
${\cal H}$ and ${\cal H}_{f}$ separate will soon become apparent.}. Then, one may 
define $V(t)$ as follows.
\begin{eqnarray}
V(t) & : & {\cal H}_{f}\rightarrow {\cal H}, \nonumber \\
A^{\mu}(t) & = & V^{\dagger}(t)A^{\mu}_{f}(t)V(t), \label{eqvtrans} \\
\psi(t) & = & V^{\dagger}(t)\psi_{f}(t)V(t).
\end{eqnarray}
Note that $V^{\dagger}(t)\neq V^{-1}(t)$. The time dependence of these relations
is shown explicitly to stress the fact that $V(t)$ can transform from free fields to interacting
fields at every instance of time. But it is {\em not} a time evolution. Also, $V(t)$ is not 
expected to be unity at any time -- not even at $t=\pm\infty$. This is because the interaction 
in terms of field operators is always present. Now, let the time evolution operator for 
interacting fields, $U(t_{1},t_{2})$ be defined as follows.
\begin{eqnarray}
U(t_{1},t_{2}) & : & {\cal H}\rightarrow {\cal H}, \nonumber \\
A^{\mu}(t_{2}) & = & U^{-1}(t_{1},t_{2})A^{\mu}(t_{1})U(t_{1},t_{2}), \\
\psi(t_{2}) & = & U^{-1}(t_{1},t_{2})\psi(t_{1})U(t_{1},t_{2}).
\end{eqnarray}
Similarly, the time evolution of the free fields can be given by the operator
$U_{f}(t_{1},t_{2})$:
\begin{eqnarray}
U_{f}(t_{1},t_{2}) & : & {\cal H}_{f}\rightarrow {\cal H}_{f}, \nonumber \\
A^{\mu}_{f}(t_{2}) & = & U^{-1}_{f}(t_{1},t_{2})A^{\mu}_{f}(t_{1})U_{f}(t_{1},t_{2}), \\
\psi_{f}(t_{2}) & = & U^{-1}_{f}(t_{1},t_{2})\psi_{f}(t_{1})U_{f}(t_{1},t_{2}).
\end{eqnarray}
\begin{figure}
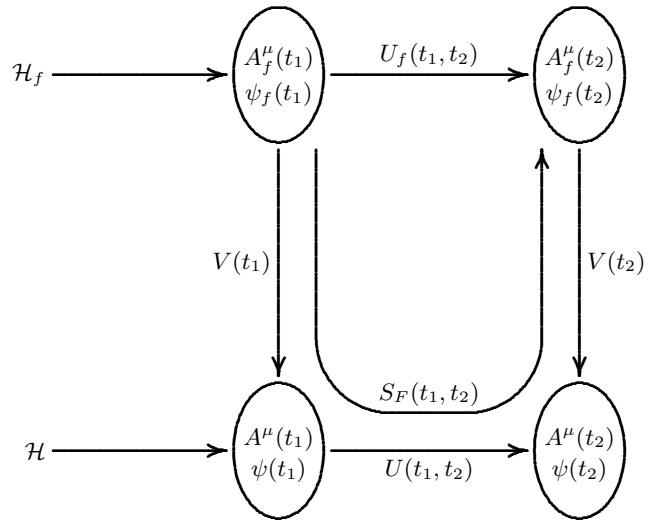

\beginpicture
\setcoordinatesystem units <1mm,1mm>
\setplotarea x from -55 to 25, y from -40 to 40
\setplotsymbol ({.})
\ellipticalarc axes ratio 2:3 360 degrees from -14 25 center at -20 25
\ellipticalarc axes ratio 2:3 360 degrees from 14 25 center at 20 25
\ellipticalarc axes ratio 2:3 360 degrees from 14 -25 center at 20 -25
\ellipticalarc axes ratio 2:3 360 degrees from -14 -25 center at -20 -25
\put {$A^{\mu}_{f}(t_{1})$} [b] at -20 25
\put {$\psi_{f}(t_{1})$} [t] at -20 24
\put {$A^{\mu}_{f}(t_{2})$} [b] at 20 25
\put {$\psi_{f}(t_{2})$} [t] at 20 24
\put {$A^{\mu}(t_{2})$} [b] at 20 -25
\put {$\psi(t_{2})$} [t] at 20 -26
\put {$A^{\mu}(t_{1})$} [b] at -20 -25
\put {$\psi(t_{1})$} [t] at -20 -26
\arrow <7pt> [.2,.67] from -50 25 to -27 25
\put {${\cal H}_{f}$} [r] at -51 25
\arrow <7pt> [.2,.67] from -50 -25 to -27 -25
\put {${\cal H}$} [r] at -51 -25
\arrow <7pt> [.2,.67] from -13 25 to 13 25
\put {$U_{f}(t_{1},t_{2})$} [b] at 0 26
\arrow <7pt> [.2,.67] from -13 -25 to 13 -25
\put {$U(t_{1},t_{2})$} [t] at 0 -26
\arrow <7pt> [.2,.67] from -20 15 to -20 -15
\put {$V(t_{1})$} [r] at -21 0
\arrow <7pt> [.2,.67] from 20 15 to 20 -15
\put {$V(t_{2})$} [l] at 21 0
\plot -15 15  -15 -10 /
\plot -5 -20  5 -20 /
\arrow <7pt> [.2,.67] from 15 -10 to 15 15
\circulararc -90 degrees from -5 -20 center at -5 -10
\circulararc 90 degrees from 5 -20 center at 5 -10
\put {$S_{F}(t_{1},t_{2})$} [b] at 0 -19
\endpicture
\caption{Mapping diagram of time evolution and interaction operators. \label{fig1}}
\end{figure}
These time evolution operators can be unitary. The relationship of $U(t_{1},t_{2})$,
$U_{f}(t_{1},t_{2})$ and $V(t)$ are shown in figure~\ref{fig1}. 
Notice that ${\cal H}$ and ${\cal H}_{f}$ are maintained as distinct. $U(t_{1},t_{2})$ is 
a mapping of ${\cal H}$ to itself and $U_{f}(t_{1},t_{2})$ is a mapping 
of ${\cal H}_{f}$ to itself. $V(t)$ is a mapping of ${\cal H}_{f}$ to ${\cal H}$. It is important to
note that $U(t_{1},t_{2})$ and $U_{f}(t_{1},t_{2})$ are bijective mappings, however,  $V(t)$ 
needs to be injective and {\em not} necessarily bijective. This is because ${\cal H}$ may be 
larger than ${\cal H}_{f}$ making some parts of ${\cal H}$ inaccessible by $V(t)$. 
To be certain if such inaccessible parts of
${\cal H}$ exist or not, one needs to construct ${\cal H}$ in a fashion similar to the Fock space
construction of ${\cal H}_{f}$. Such a construction is possible in a composite particle 
field theory (CQFT)\cite{biswas}. The operation  $S_{F}(t_{1},t_{2})$ shown in figure~\ref{fig1}
will be used in a later discussion of the scattering matrix.

It is often argued that ${\cal H}_{f}$ and ${\cal H}$ must be the same as the field operators and
their commutators are the same for both free fields and interacting fields. However, the Hilbert
space cannot be visualized as anything more than a placeholder until its basis is constructed
as a Fock space. The construction of the Fock space requires the knowledge of the equations
of motion. And the equations of motion, being different for free and interacting fields, are expected
to produce different Fock spaces.

In standard QFT, $V(t)$ is a unitary transformation from ${\cal H}_{f}$ to ${\cal H}_{f}$. This
may be acceptable at $t=\pm\infty$, but at intermediate values of $t$, it is far from the truth. As the
actual process of scattering must evolve through intermediate times, it is important to note that during
the interaction process the system will go through intermediate states that are far from being free -- in
particular, there could be temporary bound states. Loop diagrams provide approximations of
such intermediate states. However, they are bad approximations as they are constructed from
free particle states. Hence, it is not surprising that they produce infinities. A composite particle
field theory (CQFT) is able to handle such intermediate states more effectively\cite{biswas}. 
Such a model
does not need loop diagrams. The elimination of loop diagrams requires $V(t)$ to be nonunitary.
But that is seen to be just what is necessary to save perturbative QFT from Haag's theorem.

\section{Connecting the free field theory to the interacting field}
At this point, it might seem that the discussion of a free field theory might be quite unnecessary as
the physical system is always interacting and the Hilbert spaces ${\cal H}_{f}$ and
${\cal H}$ seem to be disjoint. However, the interaction picture of standard QFT has been
very successful for scattering processes without loop diagrams. And even with loop diagrams,
standard renormalization techniques have been surprisingly effective. I say surprising because
renormalization is a scheme for subtracting an infinity from another infinity to obtain a finite
quantity. Superficially seen, this ought to give different answers every time it is done -- but
it doesn't. Hence, something must be right about the standard methods. Besides, fully
nonperturbative computation for interacting fields is difficult for realistic systems. So, we need
a perturbative approach that uses the free field solutions as the starting point. Hence, we
need to find a connection between ${\cal H}_{f}$ and ${\cal H}$.

Physically speaking, the most plausible connection is obtained by assuming the vacuum
states of ${\cal H}_{f}$ and ${\cal H}$ to be identical. The single particle states should 
also be close. In fact, all possible states of ${\cal H}_{f}$ must have equivalents in
${\cal H}$ at single instances of time. These are the free n-particle states. However,
${\cal H}$ is expected to have more than just these equivalents of free n-particle states.
It can, in general, have bound states and maybe other states that cannot be visualized
as any approximation of free particles\cite{biswas}. The equivalents of the free n-particle
states in ${\cal H}$ are determined by the mapping $V(t)$ at time $t$. The image of
${\cal H}_{f}$ on ${\cal H}$ due to $V(t)$ will be called 
${\cal H}_{F}$ (${\cal H}_{F}\subset {\cal H}$).

If the vacuum state $\ket{0}$ is assumed to be the same for both ${\cal H}_{f}$ and
${\cal H}$, then
\begin{equation}
V\ket{0}=\ket{0}. \label{eq0to0}
\end{equation}
 The 1-particle states can be constructed as follows.
\begin{eqnarray}
\ket{A,1}_{f} & = & A^{\mu}_{f}\ket{0}, \label{eqketa1f} \\
\ket{A,1} & = & A^{\mu}\ket{0}, \label{eqketa1}
\end{eqnarray}
respectively for the free fields and interacting fields of radiation. The 1-particle electron
states can be written similarly as
\begin{eqnarray}
\ket{\psi,1}_{f} & = & \psi_{f}\ket{0}, \\
\ket{\psi,1} & = & \psi\ket{0},
\end{eqnarray}
Then, using equations~\ref{eqvtrans}, \ref{eq0to0}, \ref{eqketa1f} and \ref{eqketa1}, we notice that
\begin{equation}
\ket{A,1} = V^{\dagger}A^{\mu}_{f}V\ket{0}=V^{\dagger}\ket{A,1}_{f}.
\end{equation}
The adjoint of this is
\begin{equation}
\bra{A,1} =\bra{A,1}_{f}V.
\end{equation}
Hence,
\begin{equation}
\inprod{A,1}{A,1} = \bra{A,1}_{f}VV^{\dagger}\ket{A,1}_{f}.
\end{equation}
This shows that the 1-particle states of both Hilbert spaces cannot
be simultaneously normalized if $V$ is not unitary.

\section{Scattering matrix -- unitary vs. nonunitary}
The most obvious definition of the scattering matrix should be the time evolution 
$U(t_{1},t_{2})$ in ${\cal H}$ from time $t_{1}=-\infty$ to time $t_{2}=+\infty$.
This scattering matrix, $U(-\infty,+\infty)$, is clearly unitary. It describes the time evolution
of {\em all} states in ${\cal H}$ over an infinite period of time. However, standard
experiments conducted, using particle accelerators, have initial and final states that are 
free particles. The initial states are prepared to be free particles and the final states
are detected by particle detectors that are designed to detect only free particles. Hence,
experimentally, the initial and final states belong to the image ${\cal H}_{F}$ of 
${\cal H}_{f}$ in ${\cal H}$. Now, in principle, an initial free particle could develop to a final state
 that is not free. Hence,
the experimental restriction of final states also being free particles can be seen as a 
projection operation from ${\cal H}$ on to ${\cal H}_{F}$. Such a projection operation
is expected to lose probability (and energy). Hence, the experimental scattering
matrix is effectively nonunitary. For a definition of such an experimental scattering matrix,
we start with the following (see figure~\ref{fig1}).
\begin{equation}
S_{F}(t_{1},t_{2})=V^{-1}(t_{2})U(t_{1},t_{2})V(t_{1}).
\end{equation}
Here, the $V^{-1}$ operation is understood to include a projection operation from
${\cal H}$ to its subset ${\cal H}_{F}$ as $V^{-1}$ is defined only from ${\cal H}_{F}$
as domain. Clearly, $S_{F}$ is nonunitary. Now, the experimental scattering matrix can be
defined as:
\begin{equation}
S=S_{F}(-\infty,+\infty).
\end{equation}
Strictly speaking, $S$ needs to be a mapping from ${\cal H}_{F}$ to ${\cal H}_{F}$ as it
relates free particle equivalents in the interacting field Hilbert space. The operator $V$
is needed only to define ${\cal H}_{F}$ as the image of ${\cal H}_{f}$. However, here we have
defined $S$ with a $V$ and a $V^{-1}$ such that it maps ${\cal H}_{f}$ to ${\cal H}_{f}$.
This way, the operations necessary to restrict the initial and final states to free particles is implicit.
From an experimental point of view, this should not matter as ${\cal H}_{F}$ is just an
image of ${\cal H}_{f}$.

It is possible to test for such nonunitary behavior of the $S$ matrix in simplified models.
Some computations of this nature have already been done\cite{lamb}.

\section{Anatomy of a closed loop diagram}
Unfortunately, at present, there are no exact methods for computing elements of $S$.
However, reasonable approximation methods are available that do not involve
infinite renormalization. A composite particle QFT (CQFT) provides such an approximation\cite{biswas}.
Before considering such a theory, let us look at perturbative computations in standard
QFT in the light of the present definition of $S$.

\begin{figure}
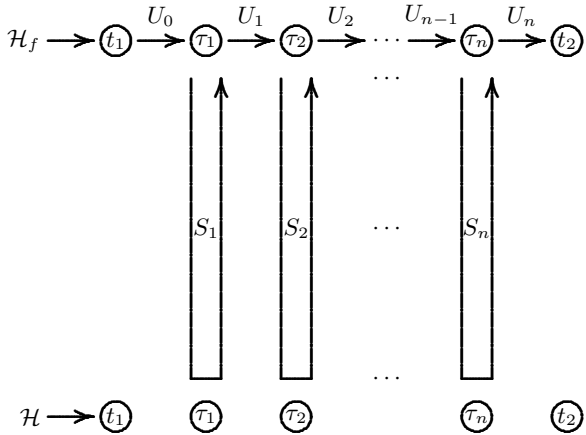

\beginpicture
\setcoordinatesystem units <1mm,1mm>
\setplotarea x from -36 to 36, y from -40 to 40
\setplotsymbol ({.})
\circulararc 360 degrees from -22 -25 center at -24 -25
\circulararc 360 degrees from -10 -25 center at -12 -25
\circulararc 360 degrees from 2 -25 center at 0 -25
\circulararc 360 degrees from 22 -25 center at 24 -25
\circulararc 360 degrees from 34 -25 center at 36 -25
\circulararc 360 degrees from -22 25 center at -24 25
\circulararc 360 degrees from -10 25 center at -12 25
\circulararc 360 degrees from 2 25 center at 0 25
\circulararc 360 degrees from 22 25 center at 24 25
\circulararc 360 degrees from 34 25 center at 36 25
\put {$t_{1}$} at -24 -25
\put {$\tau_{1}$} at -12 -25
\put {$\tau_{2}$} at 0 -25
\put {$\tau_{n}$} at 24 -25
\put {$t_{2}$} at 36 -25
\put {$t_{1}$} at -24 25
\put {$\tau_{1}$} at -12 25
\put {$\tau_{2}$} at 0 25
\put {$\tau_{n}$} at 24 25
\put {$t_{2}$} at 36 25
\arrow <7pt> [.2,.67] from -33 25 to -27 25
\put {${\cal H}_{f}$} [r] at -34 25
\arrow <7pt> [.2,.67] from -33 -25 to -27 -25
\put {${\cal H}$} [r] at -34 -25
\arrow <7pt> [.2,.67] from -21 25 to -15 25
\put {$U_{0}$} [b] at -18 27
\arrow <7pt> [.2,.67] from -9 25 to -3 25
\put {$U_{1}$} [b] at -6 27
\arrow <7pt> [.2,.67] from 3 25 to 9 25
\put {$U_{2}$} [b] at 6 27
\arrow <7pt> [.2,.67] from 15 25 to 21 25
\put {$U_{n-1}$} [b] at 18 27
\arrow <7pt> [.2,.67] from 27 25 to 33 25
\put {$U_{n}$} [b] at 30 27
\plot -14 20  -14 -20  -10 -20 /
\arrow <7pt> [.2,.67] from -10 -20 to -10 20
\put {$S_{1}$} at -12 0
\plot -2 20  -2 -20  2 -20 /
\arrow <7pt> [.2,.67] from 2 -20 to 2 20
\put {$S_{2}$} at 0 0
\plot 22 20  22 -20  26 -20 /
\arrow <7pt> [.2,.67] from 26 -20 to 26 20
\put {$S_{n}$} at 24 0
\put {$\cdots$} at 12 25
\put {$\cdots$} at 12 20
\put {$\cdots$} at 12 0
\put {$\cdots$} at 12 -20
\endpicture
\caption{Perturbative approximation of the scattering matrix in standard QFT. 
Field operators are represented by just their time argument. The notations $U_{i}$
and $S_{i}$ are defined in equation~\ref{eqsudef}. \label{fig2}}
\end{figure}

Perturbative computation in standard QFT (for a specific order) amounts to an approximation of 
$S_{F}(t_{1},t_{2})$ as a product of several operators as follows (see figure~\ref{fig2}).
\begin{eqnarray}
 S_{F}(t_{1},t_{2}) & & =  \nonumber \\
& & U_{f}(\tau_{n},t_{2})S_{F}(\tau_{n},\tau_{n}+d\tau)\times \nonumber \\
& & \times U_{f}(\tau_{n-1},\tau_{n})  \cdots \times  \nonumber \\ 
& & \times S_{F}(\tau_{2},\tau_{2}+d\tau) U_{f}(\tau_{1},\tau_{2}) \times  \nonumber \\
& & \times S_{F}(\tau_{1},\tau_{1}+d\tau)U_{f}(t_{1},\tau_{1}),
\end{eqnarray}
where $\tau_{i}$ are successive points in time and $d\tau$ is infinitesimal.
Using more compact notation, this would be as follows.
\begin{equation}
 S_{F}(t_{1},t_{2}) =  U_{n}S_{n} U_{n-1}  \cdots S_{2} U_{1} S_{1}U_{0},
\end{equation}
where, for all $i$ ($\tau_{0}=t_{1}, \tau_{n+1}=t_{2}$),
\begin{equation}
S_{i}=S_{F}(\tau_{i},\tau_{i}+d\tau),\;\;\;U_{i}=U_{f}(\tau_{i},\tau_{i+1}), \label{eqsudef}
\end{equation}
Hence, this is a propagation of a state primarily in ${\cal H}_{f}$ other than excursions of infinitesimal
durations $d\tau$ into ${\cal H}$ using $S_{F}$. These short excursions are the vertices of
Feynman diagrams. Other than at these vertices the propagation is like that of free particles\footnote{The
nonunitary nature of $S_{F}$ in the infinitesimal durations is ignored.}. As a result, intermediate bound
states are represented only by closed loops of free particle propagators between successive
vertices. But such loops produce infinities because free particles have continuous energy
spectra and the loops require integration over such spectra. A true bound state is produced by
continual binding interaction which results in a discrete energy spectrum. This approximation
using a finite number of vertices to simulate such continual interaction cannot produce
a discrete energy spectrum. Hence, the integration over all states of free particles forming
a loop gives infinities.

Composite particle QFT (CQFT) also approximates some of the interaction as happening in short intervals
of time. But it spreads the interaction center somewhat by considering each particle to have
a substructure and hence, a spatial extent. Besides, CQFT handles the time
evolution of bound states without approximating them to be collections of virtual free particles.
Hence, each bound state is treated as a single bound state instead of an integral over an
infinite number of free particle states. It is this integration over free particle states that
produces the infinities of loop diagrams. Hence, in CQFT, the propagation of a state
is considered to be in ${\cal H}$ rather than in ${\cal H}_{f}$.  Particle creation and annihilation 
still happens at the interaction centers which are extended versions of vertices. But interactions within
bound states are seen as continuous time evolution in the interacting field Hilbert space
${\cal H}$. 

In standard QFT, the propagation of a state is possible only in ${\cal H}_{f}$ and
interaction is possible only in ${\cal H}$. This is why, to generate the complete scattering matrix,
propagation has to be restricted to ${\cal H}_{f}$ and interaction has to be restricted only to 
infinitesimal excursions into ${\cal H}$. These restrictions
are eliminated in CQFT. Consequently, propagation is possible in ${\cal H}$ itself\footnote{However,
the propagators of CQFT may not be the complete set of propagators possible in ${\cal H}$.
This matter needs closer scrutiny.}. 

Constructing Feynman diagrams in CQFT is simpler than
it may appear. The tree diagrams of standard QFT are {\em all} the diagrams of CQFT. However,
it is important to note that internal lines in a diagram could represent several excited bound states.
The closed loop diagrams of standard QFT do not appear in CQFT as the propagation does not
occur in ${\cal H}_{f}$.

\end{document}